\documentclass[a4paper,12pt]{amsart}

\usepackage{amsfonts}
\usepackage{amsmath}
\usepackage{amssymb}

\usepackage{mathrsfs}
\usepackage{hyperref}

\numberwithin{equation}{section}
\theoremstyle{plain}
\newtheorem{thm}{Theorem}[section]

\theoremstyle{definition}

\newcommand{\be}{\begin{equation}}
\newcommand{\ee}{\end{equation}}

\def\R{{\mathbb R}}

\def\N{{\mathbb N^{2}}}

\def\C{{\mathbb C}}
\def\S3{{{\mathbb S}^3}}
\def\SU2{{{\rm SU}(2)}}

\def\HS{{\mathtt{HS}}}

\def\p#1{{\left({#1}\right)}}

\def\jp#1{{\left\langle{#1}\right\rangle}}

\def\Dcal{{\mathcal D}}

\DeclareMathOperator{\Tr}{Tr}

\def\Rr{{\mathbb R}}

\def\C{{\mathbb C}}

\def\R2n{{\mathbb R}^{2n}}

\def\S{{\mathcal S}}

\def\Rr{{\mathbb R}}

\def\C{{\mathbb C}}
\def\A{{\mathbb A}}

\def\H{{\mathbb H}}

\def\R2{{\mathbb R}^2}
\def\R2n{{\mathbb R}^{2n}}

\def\S{{\mathcal S}}

\def\H{{\mathcal H}}

\def\T1{{\mathbb T}^1}

\begin{document}

\title[Wave equation for 2D Landau Hamiltonian]
{Wave equation for 2D Landau Hamiltonian}

\author[Michael Ruzhansky]{Michael Ruzhansky}
\address{
  Michael Ruzhansky:
  \endgraf
  Department of Mathematics
  \endgraf
  Imperial College London
  \endgraf
  180 Queen's Gate, London, SW7 2AZ
  \endgraf
  United Kingdom
  \endgraf
  {\it E-mail address} {\rm m.ruzhansky@imperial.ac.uk}
  }
\author[Niyaz Tokmagambetov]{Niyaz Tokmagambetov}
\address{
  Niyaz Tokmagambetov:
  \endgraf
    al--Farabi Kazakh National University
  \endgraf
  71 al--Farabi ave., Almaty, 050040
  \endgraf
  Kazakhstan,
  \endgraf
   and
  \endgraf
    Department of Mathematics
  \endgraf
  Imperial College London
  \endgraf
  180 Queen's Gate, London, SW7 2AZ
  \endgraf
  United Kingdom
  \endgraf
  {\it E-mail address} {\rm n.tokmagambetov@imperial.ac.uk}
 }

\thanks{The authors were supported in parts by the EPSRC
 grant EP/K039407/1 and by the Leverhulme Grant RPG-2014-02,
 as well as by the MESRK grant 0773/GF4. No new data was collected or generated 
 during the course of research.}

\date{\today}


\begin{abstract}
This note is devoted to the study of the well-posedness of the
Cauchy problem for the Landau Hamiltonian wave equation in the
plane, with nonzero constant magnetic field. We show the well-posedness in
suitably defined Sobolev spaces taking into account the spectral
properties of the operator.
\end{abstract}

\maketitle

\section{Introduction}

The Landau Hamiltonian describing the motion of the quantum
particle in two dimensions in the constant
magnetic field is one of the classical models in quantum physics.
It was introduced and studied in late 1920s, see Landau and
Lifshitz \cite{LL77}. It turned out that the system possesses
rather unusual spectral properties: the spectrum consists of
discrete eigenvalues of infinite multiplicities lying at the points of an arithmetic
progression.
These eigenvalues are traditionally called Landau levels and the
corresponding eigenspaces are called Landau subspaces.

\smallskip

The dynamics of a particle with charge $e$ and mass $m_{\ast}$ on the Euclidean $xy$--plane interacting with a perpendicular constant homogeneous magnetic field is given by the 
Hamiltonian operator
\begin{equation} \label{eq:Hamiltonian}
\H_{0}:=\frac{1}{2m_{\ast}} \p{i h \nabla-\frac{e}{c}\A}^{2},
\end{equation}
where $h$ is the Planck's constant, $c$ is the speed of light, and
$i$ is the imaginary unit. Denoting by $2B>0$ the strength of the
magnetic field and selecting the symmetric gauge
$$
\A=-\frac{\mathrm r}{2}\times2\mathrm B=\p{- B y, B x} ,
$$
where $\mathrm r=(x, y) \in \mathbb R^{2}$, as well as setting $m_{\ast}=e=c=h=1$, 
we get the Landau Hamiltonian
\begin{equation} \label{eq:LandauHamiltonian}
\H:=\frac{1}{2} \p{\p{i\frac{\partial}{\partial x}-B
y}^{2}+\p{i\frac{\partial}{\partial y}+B x}^{2}},
\end{equation}
acting on the Hilbert space $L^{2}(\mathbb R^{2})$. It is well known 
(\cite{F28, L30}) that the
operator $\H$ has the spectrum consisting of infinite number of eigenvalues with
infinite multiplicity, of the form
\begin{equation} \label{eq:HamiltonianEigenvalues}
\lambda_{n}=\p{2n+1}B, \,\,\, n=0, 1, 2, \dots \,,
\end{equation}
and these eigenvalues are called the Euclidean Landau levels. Let us denote by $\mathcal A_{n}(\mathbb R^{2})$ the eigenspace of $\H$ corresponding to the eigenvalue $\lambda_{n}$ in \eqref{eq:HamiltonianEigenvalues}:
\begin{equation} \label{eq:HamiltonianEigenspaces}
\mathcal A_{n}(\mathbb R^{2})=\{\varphi\in L^{2}(\mathbb R^{2}), \,\, \H \varphi=\lambda_{n}\varphi\}.
\end{equation}
The following functions form an orthogonal basis for $\mathcal
A_{n}(\mathbb R^{2})$ (see \cite{ABGM15} for more details on this and the above, and also \cite{HH13}):
{\small
\begin{equation}
\label{eq:HamiltonianBasis} \left\{
\begin{split}
e^{1}_{k, n}(x,y)&=\sqrt{\frac{n!}{(n-k)!}}B^{\frac{k+1}{2}}\exp\Big(-\frac{B(x^{2}+y^{2})}{2}\Big)(x+iy)^{k}L_{n}^{(k)}(B(x^{2}+y^{2})), \,\,\, 0\leq k, {}\\
e^{2}_{j, n}(x,y)&=\sqrt{\frac{j!}{(j+n)!}}B^{\frac{n-1}{2}}\exp\Big(-\frac{B(x^{2}+y^{2})}{2}\Big)(x-iy)^{n}L_{j}^{(n)}(B(x^{2}+y^{2})), \,\,\, 0\leq j,
\end{split}
\right.
\end{equation}}
where $L^{(\alpha)}_{n}$ is the Laguerre polynomial defined as
$$
L^{(\alpha)}_{n}(t)=\sum_{k=0}^{n}(-1)^{k}C_{n+\alpha}^{n-k}\frac{t^{k}}{k!}, \,\,\, \alpha>-1.
$$
To simplify further notation we denote
\begin{equation}\label{EQ:eks}
e^{k}_{\xi}:=e^{k}_{i, n} \,\,\ \hbox{for} \,\,\,  \xi=(i,n), \,\,\, i, n=0, 1, 2, ...; \,\,\, k=1, 2.
\end{equation}

The basis \eqref{eq:HamiltonianBasis} has relations to many other subjects.
For example, see \cite{I16} for relations to the complex Hermite polynomials,
\cite{ABG12,BG14,CGG10} for relations to quantization problems,
\cite{A10} for their appearance in the time-frequency analysis, 
\cite{G08} in partial differential equations, and 
\cite{HH13} for planar
point processes. We refer to \cite{ABGM15} for further details, as well as
to a relation to Feynman and
Schwinger's work on the displacement
operator, see also Perelomov \cite[p. 35]{P86}.

We  can also refer to papers \cite{K16, N96}, where the authors investigated
properties of eigenfunctions of perturbed Hamiltonians, with descriptions of 
the asymptotics of the eigenvalues for perturbed Landau Hamiltonians in 
\cite{S14, KP04, M91, PR07, PRV13, LR14, RT08}.

In this paper, for a continuous propagation speed function $a=a(t)\geq 0$, we will be concerned with the
Cauchy problem for the Landau Hamiltonian $\H$:
\begin{equation}\label{CPa}
\left\{ \begin{split}
\partial_{t}^{2}u(t,x)-a(t)\H u(t,x)&=0, \; (t,x)\in [0,T]\times \mathbb R^{2},\\
u(0,x)&=u_{0}(x), \; x\in \mathbb R^{2}, \\
\partial_{t}u(0,x)&=u_{1}(x), \; x\in \mathbb R^{2}.
\end{split}
\right.
\end{equation}
Although the equation \eqref{CPa} is strictly hyperbolic, the coefficients of $\H$ depend on the space variables and are increasing. Therefore, although we know that the Cauchy problem
\eqref{CPa} is locally well-posed for $a\in C^{1}$, it is interesting to establish global in space properties of its solutions. This will be done in terms of the spectral properties of the operator $\H$.

The description of appearing function spaces is carried out in the spirit of \cite{DR16} using the general development of nonharmonic type analysis carried out by the authors in \cite{RT16}. The treatment of the global well-posedness in the appearing function spaces is an extension of the method developed in \cite{GR15} in the context of compact Lie groups.

\section{The main result}
\label{SEC:results}

In our results below, concerning the Cauchy problem \eqref{CPa}, we will aim at
carrying out analysis in the strictly hyperbolic case $a(t)\ge a_0>0$, $a\in C^1([0,T])$.
This is the regular time-non-degenerate case when we obtain the well-posedness
in Sobolev spaces ${H}^{s}_\H$ associated to the operator $\H$.

\begin{thm}
\label{theo_case_1}
Assume that $a\in C^1([0,T])$ and that $a(t)\ge a_0>0$.
For any $s\in\Rr$, if the Cauchy data satisfy
$(u_0,u_1)\in {H}^{1+s}_\H \times {H}^{s}_\H$,
then the Cauchy problem \eqref{CPa} has a unique solution
$u\in C([0,T],{H}^{1+s}_\H) \cap
C^1([0,T],{H}^{s}_\H)$ which satisfies the estimate
\begin{equation}
\label{case_1_last-est}
\|u(t,\cdot)\|_{{H}^{1+s}_\H}^2+\| \partial_t u(t,\cdot)\|_{{H}^s_\H}^2\leq
C (\| u_0\|_{{H}^{1+s}_\H}^2+\|u_1\|_{{H}^{s}_\H}^2).
\end{equation}
\end{thm}

\vspace{3mm}

We now define the Sobolev spaces $H^s_\H(G)$ associated to
$\H$ that appear in the theorem. Thus, for any $s\in\Rr$, we set
$$
H^s_\H(G):=\left\{ f\in\Dcal'_{\H}(\mathbb R^{2}): \H^{s/2}f\in
L^2(\mathbb R^{2})\right\},
$$
with the norm $\|f\|_{H^s_\H}:=\|\H^{s/2}f\|_{L^2}.$
The global space of distributions $\Dcal'_{\H}(\mathbb R^{2})$ is defined in
Section \ref{SEC:Prelim}.
Anticipating the material of the next section, using
Plancherel's identity \eqref{EQ:Plancherel}, in our case we can express the Sobolev norm as
\begin{equation}\label{EQ:Sob1}
\|f\|_{H^s_\H}= \p{ \sum_{\xi\in\N} (B+2B\xi_{2})^{s}
\sum_{j=1}^{2} \left| \int_{\mathbb
R^{2}}f(x)\overline{e^{j}_{\xi}(x)}dx\right|^2  }^{1/2}.
\end{equation}

\section{Fourier analysis for the Landau Hamiltonian}
\label{SEC:Prelim}

In this section we recall the necessary elements of the global Fourier analysis
that has been developed in \cite{RT16}. Although the domain $\mathbb R^{2}$
in this paper is unbounded, the following constructions carry over without any significant changes.
The space $C_{\H}^{\infty}(\mathbb R^{2}):={\rm Dom}({\H}^{\infty})$ is called the space of test functions for ${\H}$.
Here we define
$$
{\rm Dom}({\H}^{\infty}):=\bigcap_{k=1}^{\infty}{\rm Dom}({\H}^{k}),
$$
where ${\rm Dom}({\H}^{k})$ is the domain of the operator ${\H}^{k}$, in turn defined as
$$
{\rm Dom}({\H}^{k}):=\{f\in L^{2}(\mathbb R^{2}): \,\,\, {\H}^{j}f\in {\rm Dom}({\H}), \,\,\, j=0, \,1, \, 2, \ldots,
k-1\}.
$$
The Fr\'echet topology of $C_{{\H}}^{\infty}(\mathbb R^{2})$ is given by the family of norms
\begin{equation}\label{EQ:L-top}
\|\varphi\|_{C^{k}_{{\H}}}:=\max_{j\leq k}
\|{\H}^{j}\varphi\|_{L^2(\mathbb R^{2})}, \quad k\in\mathbb N_0,
\; \varphi\in C_{{\H}}^{\infty}(\mathbb R^{2}).
\end{equation}

The space $$\mathcal D'_{{\H}}(\mathbb R^{2}):=\mathcal L(C_{\H}^{\infty}(\mathbb R^{2}),
\mathbb C)$$ of linear continuous functionals on
$C_{\H}^{\infty}(\mathbb R^{2})$ is called the space of
${\H}$-distributions.
We can understand the continuity here in terms of the topology
\eqref{EQ:L-top}.
For
$w\in\mathcal D'_{{\H}}(\mathbb R^{2})$ and $\varphi\in C_{\H}^{\infty}(\mathbb R^{2})$,
we shall write
$$
w(\varphi)=\langle w, \varphi\rangle.
$$
For any $\psi\in C_{{\H}}^{\infty}(\mathbb R^{2})$, the functional
$$
C_{\H}^{\infty}(\mathbb R^{2})\ni \varphi\mapsto\int_{\mathbb R^{2}}{\psi(x)} \, \varphi(x)\, dx
$$
is an ${\H}$-distribution, which gives an embedding $\psi\in
C_{{\H}}^{\infty}(\mathbb R^{2})\hookrightarrow\mathcal D'_{\H}(\mathbb R^{2})$.

Taking into account the fact that the eigenfunctions of the Landau Hamiltonian in \eqref{eq:HamiltonianBasis} come in pairs, it will be convenient to group them together in the way
suggested by the notation \eqref{EQ:eks}. This leads to the following definitions.
Let $\mathcal S(\N)$ denote the space of rapidly decaying
functions $\varphi:\N\rightarrow\mathbb C^{2\times2}$ of the form
$$
\varphi:=
\p{\begin{matrix} \varphi_{11}  & 0 {}\\
0 &\varphi_{22}
\end{matrix}}.
$$

That is,
$\varphi\in\mathcal S(\N)$ if for any $M<\infty$ there
exists a constant $C_{\varphi, M}$ such that
$$
|\varphi(\xi)|\leq C_{\varphi, M}\langle\xi\rangle^{-M}
$$
holds for all $\xi\in\N$, where we denote
$$\langle\xi\rangle:=|\sqrt{\lambda_{\xi_{2}}}|=|\sqrt{(2\xi_{2}+1)B}|.$$

The topology on $\mathcal
S(\N)$ is given by the seminorms $p_{k}$, where
$k\in\mathbb N_{0}$ and
$$
p_{k}(\varphi):=\sup_{\xi\in\N}\langle\xi\rangle^{k}|\varphi(\xi)|.
$$

We now define the $\H$-Fourier transform on $C_{\H}^{\infty}(\mathbb R^{2})$ as the mapping
$$
(\mathcal F_{\H}f)(\xi)=(f\mapsto\widehat{f}):
C_{\H}^{\infty}(\mathbb R^{2})\rightarrow\mathcal S(\N)
$$
by the formula
\begin{equation}
\label{FourierTr}
\widehat{f}(\xi):=(\mathcal F_{\H}f)(\xi)=\int_{\mathbb R^{2}}f(x)\overline{e_{\xi}(x)}dx,
\end{equation}
where
$$
e_{\xi}(x)=\p{\begin{matrix} e^{1}_{\xi}(x)  & 0 {}\\
0 & e^{2}_{\xi}(x)
\end{matrix}}.
$$

The $\H$-Fourier transform
$\mathcal F_{\H}$ is a bijective homeomorphism from $C_{{\H}}^{\infty}(\mathbb R^{2})$ to $\mathcal S(\N)$.
Its inverse  $$\mathcal F_{\H}^{-1}: \mathcal S(\N)
\rightarrow C_{\H}^{\infty}(\mathbb R^{2})$$ is given by
\begin{equation}
\label{InvFourierTr} (\mathcal F^{-1}_{{\H}}h)(x)=\sum_{\xi \in \N}\Tr\p{h(\xi)e_{\xi}(x)},\quad h\in\mathcal S(\N),
\end{equation}
so that the Fourier inversion formula becomes
\begin{equation}
\label{InvFourierTr0}
f(x)=\sum_{\xi\in\N}\Tr\p{\widehat{f}(\xi)e_{\xi}(x)}
\quad \textrm{ for all } f\in C_{{\H}}^{\infty}(\mathbb R^{2}).
\end{equation}

The Plancherel's identity taking the form
\begin{equation}\label{EQ:Plancherel}
\|f\|_{L^{2}(\mathbb R^{2})}=\p{\sum_{\xi\in\N}
\|\widehat{f}(\xi)\|_{\HS}^{2}}^{1/2}=:
\|\widehat{f}\|_{\ell^{2}(\N)},
\end{equation}
which we take as the definition of the norm on the Hilbert space
$\ell^{2}(\N)$, and where
$ \|\widehat{f}(\xi)\|_{\HS}^{2}=\Tr(\widehat{f}(\xi)\overline{\widehat{f}(\xi))}$ is the
Hilbert--Schmidt norm of the matrix $\widehat{f}(\xi)$.

One can readily check that test functions and distributions on $\mathbb R^{2}$ can be characterised in terms of
their Fourier coefficients. Thus, we have
$$
f\in C^{\infty}_{\H}(\mathbb R^{2})\Longleftrightarrow
\forall N \;\exists C_{N} \textrm{ such that }
\|\widehat{f}(\xi)\|_{\HS}\leq C_{N} \jp{\xi}^{-N}
\textrm{ for all } \xi\in\N.
$$
Also, for distributions, we have
$$
u\in \Dcal'_{\H}(\mathbb R^{2})
\Longleftrightarrow
\exists M \;\exists C \textrm{ such that }
\|\widehat{u}(\xi)\|_{\HS} \leq C\jp{\xi}^{M}
\textrm{ for all } \xi\in\N .
$$

In general, given a linear continuous operator $L:C^{\infty}_{\H}(\mathbb R^{2})\to C^{\infty}_{\H}(\mathbb R^{2})$
(or even $L:C^{\infty}_{\H}(\mathbb R^{2})\to \Dcal'_{\H}(\mathbb R^{2})$), we can define its matrix symbol by
$\sigma_{L}(x,\xi):=e_{\xi}(x)^{-1} (L e_\xi)(x)\in \C^{2\times2}$, where
$L e_\xi$ means that we apply $L$ to the matrix components of $e_\xi(x)$, provided that
$e_{\xi}(x)$ is invertible in a suitable sense.
In this case we may prove that
\begin{equation}\label{EQ:T-op}
Lf(x)=\sum_{\xi\in\N} \Tr\p{e_\xi(x)\sigma_{L}(x,\xi)\widehat{f}(\xi)}.
\end{equation}
The correspondence between operators and symbols is one-to-one.
The quantization \eqref{EQ:T-op} has been extensively studied in
\cite{Ruzhansky-Turunen:BOOK,Ruzhansky-Turunen:IMRN} in the setting of compact Lie groups, and in \cite{RT16} in the setting of (non-self-adjoint) boundary value problems, to which we
may refer for its properties and for the corresponding symbolic calculus.
However, the situation with the Landau Hamiltonian is essentially simpler since this operator can be treated as an `invariant' operator in the corresponding global calculus.

The operator $\H$ is formally self-adjoint, therefore its symbol
$\sigma_\H(\xi)$ is independent of $x$ and can be diagonalised by a choice of the basis in the in $\mathbb R^{2}$. Moreover, the operator $\H$ is positive. Therefore, without loss of generality, we can always write
\begin{equation}\label{EQ:subL-symbol}
\sigma_{\H}(\xi)=
\left(\begin{matrix}
\nu_1^2(\xi) &  0\\
0  & \nu_2^2(\xi)
\end{matrix}\right),
\end{equation}
for some $\nu_j(\xi)\geq 0$. Indeed, we have
$\nu_{j}^{2}(\xi)=B(1+2\xi_{2})$ for $j=1,2$.

Consequently, we can also define Sobolev spaces $H^s_\H(G)$ associated to
$\H$. Thus, for any $s\in\Rr$, we set
\begin{equation}\label{EQ:HsL}
H^s_\H(G):=\left\{ f\in\Dcal'_{\H}(\mathbb R^{2}): \H^{s/2}f\in
L^2(\mathbb R^{2})\right\},
\end{equation}
with the norm $\|f\|_{H^s_\H}:=\|\H^{s/2}f\|_{L^2}.$ Using
Plancherel's identity \eqref{EQ:Plancherel}, we can write
\begin{multline}\label{EQ:Hsub-norm}
\|f\|_{H^s_\H}=\|\H^{s/2}f\|_{L^2}=
\p{\sum_{\xi\in\N} \|\sigma_\H(\xi)^{s/2}\widehat{f}(\xi)\|_\HS^2}^{1/2} \\
= \p{ \sum_{\xi\in\N}  (B+2B\xi_{2})^{s} \sum_{j=1}^{2}
|\widehat{f}(\xi)_{jj}|^2  }^{1/2}= \p{ \sum_{\xi\in\N}
(B+2B\xi_{2})^{s} \sum_{j=1}^{2} \left| \int_{\mathbb
R^{2}}f(x)\overline{e^{j}_{\xi}(x)}dx\right|^2 }^{1/2},
\end{multline}
justifying the expression \eqref{EQ:Sob1}.

\section{Proof of the theorem}
\label{SEC:reduction}

The operator $\H$  has the symbol \eqref{EQ:subL-symbol}, which we can write
in matrix components as
$$
\sigma_{\H}(\xi)_{mk}=B(1+2\xi_{2})\delta_{mk}, \; 1\leq m,k\leq
2,
$$
with $\delta_{mk}$ standing for the Kronecker's delta.
Taking the $\H$-Fourier transform of
\eqref{CPa}, we obtain the collection of Cauchy problems for
matrix-valued Fourier coefficients:
\begin{equation}\label{CPa-FC}
\partial_{t}^{2}\widehat{u}(t,\xi)-a(t)\sigma_{\H}(\xi)\widehat{u}(t,\xi)=0,
\; \xi\in\N.
\end{equation}
Writing this in the matrix form, we see that this is equivalent to the system
$$
\partial_{t}^{2} \widehat{u}(t,\xi)+
a(t) \left(\begin{matrix}
B(1+2\xi_{2}) &  0 \\
0  &  B(1+2\xi_{2})
\end{matrix}\right) \widehat{u}(t,\xi)=0,
$$
where we put explicitly the diagonal symbol
$\sigma_{\H}(\xi)$.
Rewriting \eqref{CPa-FC} in terms of matrix coefficients
$\widehat{u}(t,\xi)=\left(\widehat{u}(t,\xi)_{mk}\right)_{1\leq m,k\leq 2}$,
we get the equations
\begin{equation}\label{EQ:WE-scalars}
\partial_{t}^{2} \widehat{u}(t,\xi)_{mk}+ a(t) B(1+2\xi_{2})
 \widehat{u}(t,\xi)_{mk}=0,\qquad \xi\in\N,\;
1\leq m,k\leq 2.
\end{equation}
The main point of our further analysis is that we can make an individual
treatment of the equations in \eqref{EQ:WE-scalars}.
Thus, let us fix $\xi\in\N$ and $m,k$ with $1\leq m,k\leq 2$,
and let us denote
$\widehat{v}(t,\xi):=\widehat{u}(t,\xi)_{mk}$.
We then study the Cauchy problem
\begin{equation}\label{EQ:WE-v}
\partial_{t}^{2} \widehat{v}(t,\xi)+ a(t) B(1+2\xi_{2})
 \widehat{v}(t,\xi)=0,\quad
 \widehat{v}(t,\xi)=\widehat{v}_{0}(\xi), \;
 \partial_{t}\widehat{v}(t,\xi)=\widehat{v}_{1}(\xi),
\end{equation}
with $\xi,m$ being parameters, and want to derive estimates
for $\widehat{v}(t,\xi)$. Combined with the characterisation \eqref{EQ:Hsub-norm} of
Sobolevspaces this will yield the well-posedness
results for the original Cauchy problem \eqref{CPa}.

In the sequel, for fixed $m$, we set
\begin{equation}\label{xi_l}
|\xi|_{\nu}^{2}:=B(1+2\xi_{2}).
\end{equation}
Hence, the equation in \eqref{EQ:WE-v} can be written as
\begin{equation}
\label{eq_xi}
\partial_{t}^{2} \widehat{v}(t,\xi)+a(t)|\xi|_\nu^2\widehat{v}(t,\xi)=0.
\end{equation}
We now proceed
with a standard reduction to a first order system of this equation and define the corresponding energy.
The energy estimates will be given in terms of $t$ and $|\xi|_\nu$ and we then go back to $t$,
$\xi$ and $m$ by using \eqref{xi_l}.

We can now do the natural energy construction for \eqref{eq_xi}. We use the transformation
\[
\begin{split}
V_1&:=i|\xi|_\nu\widehat{v},\\
V_2&:= \partial_t \widehat{v}.
\end{split}
\]
It follows that the equation \eqref{eq_xi} can be written as the first order system
\begin{equation}\label{EQ:system}
\partial_t V(t,\xi)=i|\xi|_\nu A(t)V(t,\xi),
\end{equation}
where $V$ is the column vector with entries $V_1$ and $V_2$ and
\[
A(t)=\left(
    \begin{array}{cc}
      0 & 1\\
      a(t) & 0 \\
           \end{array}
  \right).
\]
The initial conditions $\widehat{v}(0,\xi)=\widehat{v}_{0}(\xi)$, $\partial_{t}\widehat{v}(0,\xi)=\widehat{v}_{1}(\xi)$
are transformed into
\[
V(0,\xi)=\left(
    \begin{array}{c}
      i|\xi|_\nu \widehat{v}_0(\xi)\\
      \widehat{v}_{1}(\xi)
     \end{array}
  \right).
\]
Note that the matrix $A$ has eigenvalues $\pm\sqrt{a(t)}$ and symmetriser
\[
S(t)=\left(
    \begin{array}{cc}
      2a(t) & 0\\
      0 & 2 \\
           \end{array}
  \right).
\]
By definition of the symmetriser we have that
\[
SA-A^\ast S=0.
\]
It is immediate to prove that
\begin{equation}
\label{est_sym}
2\min_{t\in[0,T]}(a(t),1)|V|^2\le (SV,V)\le 2\max_{t\in[0,T]}(a(t),1)|V|^2,
\end{equation}
where $(\cdot,\cdot)$ and $|\cdot|$ denote the inner product and the norm in $\C$, respectively.


Since $a(t)>0$, $a\in C^1([0,T])$, it is clear that there exist constants $a_0>0$ and $a_1>0$ such that
\[
a_0=\min_{t\in[0,T]}a(t)
\; \textrm{ and } \;
a_1=\max_{t\in[0,T]}{a(t)}.
\]
Hence \eqref{est_sym} implies,
\begin{equation}
\label{est_sym_1}
c_0|V|^2=2\min(a_0,1)|V|^2\le (SV,V)\le 2\max(a_1,1)|V|^2=c_1|V|^2,
\end{equation}
with $c_0,c_1>0$.
We then define the energy
$$E(t,\xi):=(S(t)V(t,\xi),V(t,\xi)).$$
We get, from \eqref{est_sym_1}, that
\begin{align*}
\partial_t E(t,\xi)&=(\partial_t S(t)V(t,\xi),V(t,\xi))+(S(t)\partial_t V(t,\xi),V(t,\xi))\\
&\,\,\,\,\,\,\,\,\,\,\,\,\,\,\,\,\,\,\,\,\,\,\,\,\,\,\,\,\,\,\,\,\,\,\,\,\,\,\,\,\,\,\,\,\,\,\,\,\,\,\,\,\,\,\,\,\,\,\,\,\,\,\,\,
+(S(t)V(t,\xi),\partial_t V(t,\xi))\\
&=(\partial_t S(t)V(t,\xi),V(t,\xi))+i|\xi|_\nu (S(t)A(t)V(t,\xi),V(t,\xi))\\
&\,\,\,\,\,\,\,\,\,\,\,\,\,\,\,\,\,\,\,\,\,\,\,\,\,\,\,\,\,\,\,\,\,\,\,\,\,\,\,\,\,\,\,\,\,\,\,\,\,\,\,\,\,\,\,\,\,\,\,\,\,\,\,\,
-i|\xi|_\nu (S(t)V(t,\xi),A(t)V(t,\xi))\\
&=(\partial_t S(t)V(t,\xi),V(t,\xi))+i|\xi|_\nu ((SA-A^\ast S)(t)V(t,\xi),V(t,\xi))\\
&=(\partial_t S(t)V(t,\xi),V(t,\xi))\\
&\le \Vert \partial_t S\Vert |V(t,\xi)|^2\\
&\le c' E(t,\xi)
\end{align*}
i.e. we obtain
\begin{equation}
\label{E_1}
\partial_t E(t,\xi)\le c' E(t,\xi),
\end{equation}
for some constant $c'>0$. By Gronwall's lemma applied to inequality
\eqref{E_1} we conclude that for all $T>0$ there exists $c>0$ such that
\[
E(t,\xi)\le c E(0,\xi).
\]
Hence, inequalities \eqref{est_sym_1} yield
\[
c_0|V(t,\xi)|^2\le E(t,\xi)\le c E(0,\xi)\le cc_1|V(0,\xi)|^2,
\]
for constants independent of $t\in[0,T]$ and $\xi$. This allows us to write the following statement:
there exists a constant $C_1>0$ such that
\begin{equation}
\label{case_1_est}
|V(t,\xi)|\le C_1 |V(0,\xi)|,
\end{equation}
for all $t\in[0,T]$ and $\xi$. Hence
\[
|\xi|_\nu^2 |\widehat{v}(t,\xi)|^2+|\partial_t \widehat{v}(t,\xi)|^2
\le C_1'( |\xi|_\nu^2 |\widehat{v}_0(\xi)|^2+|\widehat{v}_1(\xi)|^2).
\]
Recalling the notation
$\widehat{v}(t,\xi)=\widehat{u}(t,\xi)_{mk}$ and
$|\xi|_\nu=B(1+2\xi_{2})$, this means
\begin{equation}
\label{case_1_est_mn} B(1+2\xi_{2})
|\widehat{u}(t,\xi)_{mk}|^2+|\partial_t \widehat{u}(t,\xi)_{mk}|^2
\le C_1'( B(1+2\xi_{2})
|\widehat{u}_0(\xi)_{mk}|^2+|\widehat{u}_1(\xi)_{mk}|^2)
\end{equation}
for all $t\in[0,T]$, $\xi\in\N$ and $1\le m,k\le 2$, with the
constant $C_1'$ independent of $\xi$, $m,k$.
Now we recall that by Plancherel's equality, we have
$$
\|\partial_t u(t,\cdot)\|_{L^2}^2=\sum_{\xi\in\N}\|\partial_t \widehat{u}(t,\xi)\|_\HS^2=
\sum_{\xi\in\N} \sum_{m,k=1}^2 |\partial_t \widehat{u}(t,\xi)_{mk}|^2
$$
and
$$
\|\H u(t,\cdot)\|_{L^2}^2=\sum_{\xi\in\N}   \| \sigma_\H(\xi)
\widehat{u}(t,\xi)\|_\HS^2= \sum_{\xi\in\N} \sum_{m,k=1}^2
B(1+2\xi_{2}) |\widehat{u}(t,\xi)_{mk}|^2.
$$
Hence, the estimate \eqref{case_1_est_mn} implies that
\begin{equation}
\label{case_1_last}
\|\H u(t,\cdot)\|_{L^2}^2+\|\partial_t u(t,\cdot)\|_{L^2}^2\leq
C (\|\H u_0\|_{L^2}^2+\|u_1\|_{L^2}^2),
\end{equation}
where the constant $C>0$ does not depend on $t\in[0,T]$. More
generally, multiplying \eqref{case_1_est_mn} by powers of
$(B+2B\xi_{2})$, for any $s$, we get
\begin{multline}
\label{case_1_est_mn2} (B+2B\xi_{2})^{1+s}
|\widehat{u}(t,\xi)_{mk}|^2+(B+2B\xi_{2})^{s}  |\partial_t
\widehat{u}(t,\xi)_{mk}|^2 \\
\le C_1'( (B+2B\xi_{2})^{1+s}
|\widehat{u}_0(\xi)_{mk}|^2+(B+2B\xi_{2})^{s}
|\widehat{u}_1(\xi)_{mk}|^2).
\end{multline}
Taking the sum over $\xi$, $m$ and $k$ as above, this yields
the estimate \eqref{case_1_last-est}.

\end{document}